\documentclass[english,aps,nofootinbib,eqsecnum,superscriptaddress]{revtex4-1}
\usepackage[T1]{fontenc}
\usepackage[utf8]{luainputenc}
\usepackage{amsmath}
\usepackage{amssymb}
\usepackage{graphicx}
\usepackage{esint}
\usepackage{lipsum}
\usepackage{xcolor}

\makeatletter

\numberwithin{figure}{section}

\usepackage{array}
\usepackage{appendix}
\usepackage{multirow}

\allowdisplaybreaks[4]

\usepackage{float}
\usepackage{wasysym}
\usepackage{etoolbox}

\usepackage{titlesec}
\usepackage[marginal]{footmisc}
\usepackage{braket}

\usepackage{verbatim}
\usepackage{bm}
\usepackage{babel}
\numberwithin{equation}{section}
\DeclareMathSizes{12}{12}{7}{7}
\renewcommand{\theequation}{\arabic{section}.\arabic{equation}}

\renewcommand\thesubsection{\thesection.\arabic{subsection}}

\renewcommand{\baselinestretch}{1.2}
\@booleantrue\preprintsty@sw
\def\frontmatter@abstractheading{\centerline{\textbf{Abstract}}}
\g@addto@macro\abstract{\ignorespaces}

\makeatother

\begin{document}

\title{
Describing Migdal effects in diamond crystal with atom-centered\\ localized
Wannier functions}

\author{Zheng-Liang Liang}
\email{liangzl@mail.buct.edu.cn}
\affiliation{College of Mathematics and Physics, Beijing University of Chemical Technology\\Beijing 100029, China}

\author{Lin Zhang}
\email{zhanglin@ucas.ac.cn}
\affiliation{School of Nuclear Science and Technology, University of Chinese Academy of Sciences,\\Beijing, 100049, China}

\author{Fawei Zheng}
\email{zheng_fawei@iapcm.ac.cn}
\affiliation{Institute of Applied Physics and Computational Mathematics\\Beijing, 100088, China}

\author{Ping Zhang}
\email{zhang_ping@iapcm.ac.cn}
\affiliation{Institute of Applied Physics and Computational Mathematics\\Beijing, 100088, China}
\affiliation{School of Physics and Physical Engineering, Qufu Normal University\\ Qufu, 273165, China}

\begin{abstract}
 Recent studies have theoretically investigated the atomic excitation
and ionization induced by the dark matter (DM)-nucleus scattering,
and it is found that the suddenly recoiled atom is much more likely
to excite or lose its electrons than expected. Such  phenomenon is  called
the ``Migdal effect''. In this paper, we extend the established
strategy to describe the Migdal effect in isolated atoms to the case
in semiconductors under the framework of tight-binding (TB) approximation.
Since the localized aspects of electrons are respected in form of
the Wannier functions (WFs), the extension of the existing Migdal
approach for isolated atoms is much more natural, while the extensive
nature of electrons in solids is reflected in the hopping integrals.
We take diamond target as a concrete  proof of principle
for the methodology, and calculate relevant energy spectra and projected
sensitivity of such diamond detector. It turns out that our method
as a preliminary attempt is theoretically self-consistent and  practically effective.
\end{abstract}
\maketitle

\section{Introduction}

The identity of the dark matter (DM) is one of the most puzzling problems
in modern physics. Although there has been overwhelming evidence for
its existence from astrophysics and cosmology, its nature still remains
a mystery from a particle physical perspective. In decades, tremendous
efforts are invested into the search for the weakly interacting massive
particles (WIMPs), which not only take root in mere theoretical motivations, but also naturally explain the observed
relic abundance in the context of thermal freeze-out. Owing to the
spectacular improvements in sensitivity over recent years, the frontier
of the detection has been pushed to the DM mass range around the sub-GeV
scale, where traditional experiments (\textit{e.g.,} XENON1T~\cite{Aprile:2018dbl,Aprile:2019xxb},
LUX~\cite{Akerib:2018hck}, PandaX~\cite{Ren:2018gyx}, \textit{etc}.~\cite{Agnese:2017jvy,Liu:2019kzq,Abdelhameed:2019hmk,Agnes:2018ves,Armengaud:2019kfj})
are expected to turn insensitive. This has motivated alternative proposals
based on new detection channels and  targets, such as  with semiconductors~\cite{Essig:2011nj,Graham:2012su,
Essig:2015cda,Hochberg:2016sqx}, superconductors~\cite{Hochberg:2015pha,Hochberg:2016ajh}, Dirac materials~\cite{Hochberg:2017wce,Coskuner:2019odd,Geilhufe:2019ndy}, superfluid helium~\cite{Knapen:2016cue,Caputo:2019cyg,Caputo:2019xum}, and through phonon excitations~\cite{Knapen:2017ekk,Griffin:2018bjn,Campbell-Deem:2019hdx}, as well as other proposals and analyses~\cite{Essig:2012yx,Lee:2015qva,Hochberg:2015fth,
Bloch:2016sjj,Derenzo:2016fse,Hochberg:2016ntt,Essig:2016crl,
Kadribasic:2017obi,Essig:2017kqs,Arvanitaki:2017nhi,
Budnik:2017sbu,SHARMA2017326,Cavoto:2017otc,Liang:2018bdb,Heikinheimo:2019lwg,
Trickle:2019nya,Trickle:2019ovy,Catena:2019gfa}.

However, a recent research~\cite{Ibe2018} clarified that the sensitivity
of these conventional strategies have been significantly underestimated.
In contrast to the previous impression that electrons are so tightly bound to
the atom that the sudden boost of a recoiled atom cannot ``shake
off''  the outer electrons, the authors of Ref.~\cite{Ibe2018}
pointed out that in realistic case it takes some time for the electrons
to catch up with the struck atom, and excitation and ionization are
found to be more frequent than anticipated. Such phenomenon is termed
``Migdal effect'' in the DM literature~\cite{Bernabei:2007jz,Dolan:2017xbu,Ibe2018,Bell:2019egg,Baxter:2019pnz,Emken:2019tni,Essig:2019xkx}.

The purpose of this study is to extend the formalism developed for
isolated atoms in Ref.~\cite{Ibe2018} to the case of semiconductors\footnote[1]{\renewcommand{\baselinestretch}{1}\selectfont  Ref.~\cite{Essig:2019xkx} first investigated the Migdal effect in semiconductor targets by exploring the connection between the DM-electron scattering in semiconductors and  Migdal processes in isolated atoms.}.
Unlike the electrons exclusively bound to an individual atom, the
delocalized electrons in solids are free to hop between neighboring
ions, which makes a direct application of the method developed in
Ref.~\cite{Ibe2018} to the crystalline environment rather dubious.
On the one hand, if one follows the rest frame of the suddenly recoiled nucleus in solids,
the rest of the nuclei in solids will no longer be stationary either. On the other
hand, if one follows the whole recoiled crystal, the highly local
impulsive effect caused by the incident DM particle cannot be appropriately
accounted for. To pursue a reasonable extension, we resort to
the tight-binding (TB) approximation in which both the local and extensive
characteristics of the Migdal effect in crystal are taken into consideration
simultaneously. To be specific, we first re-express the Bloch wavefunctions
in terms of the Wannier functions (WFs) that reflect the localized
aspects of the itinerant electrons, and then we impose the Galilean
transformation operator exclusively onto those WFs associated with
the struck atom, and as a consequence the extensive aspects of electrons
are effectively encoded in the hopping integrals. As a natural representation
of localized orbital for extended systems, WFs play a key role
 in various applications related to  local phenomena, such as defects,
excitons,  electronic polarization and magnetization,
as well as in formal discussions of the  Hubbard models of strongly
correlated systems~\cite{RevModPhys.84.1419,doi:10.1002/9781119148739.ch6}.
Dealing with the Migdal effect in crystalline solids is another interesting
application of the WFs.

To deal with the electronic excitation rate via recoiled ions, the
quantization of vibration seems to be an alternative approach, since
related studies and applicable tools have existed for long in areas
such as neutron scattering in solid state (for a review, see~\cite{Schober2014}),
and recently in DM detection~\cite{Schutz:2016tid,Knapen:2017ekk,Griffin:2018bjn,Acanfora:2019con,Campbell-Deem:2019hdx}.
However, taking a look at the Feynman rules of the phonon excitation process,
one finds that each phonon external leg contributes a term proportional
to $\boldsymbol{\epsilon}_{\mathbf{k},\alpha}^{*}\cdot\mathbf{q}/\sqrt{m_{N}\,\omega_{\mathbf{k},\alpha}}$,
where $\mathbf{q}$ is the transferred momentum,  $\boldsymbol{\epsilon}_{\mathbf{k},\alpha}$ and $\omega_{\mathbf{k},\alpha}$
are the phonon eigenvector and the eigen-frequency of branch $\alpha$
at momentum  $\mathbf{k}$, respectively, and $m_{N}$ the
mass of nucleus. Therefore, in the DM mass range above a few MeV,
multi-phonon effects are no longer negligible, and if a large momentum
transfer is involved, all kinetically possible processes have to be
taken into account, making the problem seemingly intractable. However,
it is found through an isotropic harmonic oscillator model that
in the limit $q=\left|\mathbf{q}\right|\rightarrow\infty$, the effects
of all the multi-phonon terms can be well summarized with the impulse
approximation~\cite{Schober2014}, where all the vibrational effects
are encoded in a free recoiled atom in a very short timescale. It
is during this period of time, excitation occurs. So, in the DM mass
range of sub-GeV, the impulse approximation, or the nuclear recoil
interpretation, is the appropriate approach to depict the Migdal
effect in solids.

In discussion we take diamond crystal as a concrete example to demonstrate
the feasibility of the TB approach to describe the Migdal effect in
solid detectors. Diamond  is a promising material
for  DM detection, possessing numerous
advantages over traditional silicon and germanium semiconductor detectors,
such as the lighter mass of carbon nucleus that brings about a lower
DM mass threshold, the long-lived and hard phonon modes that facilitate
the phonon collection, and the ability to withstand strong electric fields
that drive the ionized electrons across the bulk material, \textit{etc}.~\cite{Kurinsky:2019pgb}.

This paper is organized as follows. In Sec.~\ref{sec:Sec_TB}, we
first take a brief review of the Migdal effect in atoms, and then
outline the TB framework to describe the Migdal effect in crystalline
solids. In Sec.~\ref{sec:WFs in Migdal Effects}, we put into practice
the TB approach by use of \textit{ab initio} density functional theory~(DFT) code $\mathtt{Quantum\,Espresso}$~\cite{Giannozzi_2009} and WF-generation tool $\mathtt{Wannier\,90}$,
concretely calculating the Migdal excitation event rate and relevant
energy spectrum for crystalline diamond. Conclusion and open discussions
are arranged in Sec.~\ref{sec:Conclusions}.

Throughout the paper the natural units $\hbar=c=1$ is adopted, while
velocities are expressed with units of $\mathrm{km/s}$ in text for
convenience.

\section{\label{sec:Sec_TB}Electronic excitation in the tight-binding description}

We begin this section with a short review of the treatment of the
Migdal effect in isolated atoms, and then generalize its application
to the electronic bands in crystalline solids.

\subsection{Migdal effect in isolated atoms}

In an atom, the excitation/ionization of electrons can be reasonably
estimated by using the Migdal's approach~\cite{Ibe2018}, in which
the excitation/ionization is a dynamical consequence
of  suddenly moving electrons in the rest
frame of a recoiled nucleus. To account for the Migdal effect in the
rest frame of the struck nucleus we invoke the Galilean boost
operator $e^{im_{e}\mathbf{v}\cdot\hat{\mathbf{r}}}$ . For the given
velocity $\mathbf{v}$, the operator $e^{im_{e}\mathbf{v}\cdot\hat{\mathbf{r}}}$
boosts the electron state at rest ($m_{e}$ being the electron mass
and $\hat{\mathbf{r}}$ the electron position operator) to the inertial
frame moving with velocity $\mathbf{v}$. To see this, assuming $\ket{\mathbf{p}}$
is the eigenstate of the momentum operator $\hat{\mathbf{p}}$ with
eigen momentum $\mathbf{p}$, it is straightforward to verify that
\begin{eqnarray}
\hat{\mathbf{p}}\left(e^{im_{e}\mathbf{v}\cdot\hat{\mathbf{r}}}\ket{\mathbf{p}}\right) & = & \left(\mathbf{p}+m_{e}\mathbf{v}\right)\,e^{im_{e}\mathbf{v}\cdot\hat{\mathbf{r}}}\ket{\mathbf{p}}.
\end{eqnarray}

Thus, keeping pace with the struck nucleus, one can schematically
express the excitation/ionization probability as
\begin{eqnarray}
\mathcal{P} & \propto & \left|\left\langle \psi_{2}\right|e^{i\mathbf{q}_{e}\cdot\hat{\mathbf{r}}}\left|\psi_{1}\right\rangle \right|^{2},
\end{eqnarray}
where $\left|\psi_{1}\right\rangle $ and $\left|\psi_{2}\right\rangle $
represent the initial and final states, respectively, sandwiching
the Galilean transformation operator $e^{i\mathbf{q}_{e}\cdot\hat{\mathbf{r}}}$
that boosts the bound electron in the opposite direction to the recoiled
nucleus with Galilean momentum $\mathbf{q}_{e}\equiv\left(m_{e}/m_{N}\right)\mathbf{q}$,
with the nucleus mass $m_{N}$, and the DM transferred momentum $\mathbf{q}=\mathbf{p}_{\chi,f}-\mathbf{p}_{\chi,i}$,
with $\mathbf{p}_{\chi,i}$ and $\mathbf{p}_{\chi,f}$ being the DM
momenta before and after the scattering respectively. To compute the
excitation/ionization rate, one also needs to integrate over  momentum
transfer $\mathbf{q}$ and  DM velocity $\mathbf{w}$. As a result,
the transition event rate for a DM particle with incident velocity
$\mathbf{w}$ to excite a bound electron via the Migdal process from
level 1 to level 2 can be expressed in the following form\footnote[2]{\renewcommand{\baselinestretch}{1}\selectfont Here we omit the nuclear form factor since the transfer momentum is so small for sub-GeV DM that  the structure of nucleus is irrelevant for a coherent DM-nucleus scattering.}:
\begin{eqnarray}
R_{1\rightarrow2} & = & \frac{\rho_{\chi}}{m_{\chi}}\left\langle \sigma_{1\rightarrow2}w\right\rangle \nonumber \\
 & = & \frac{\rho_{\chi}}{m_{\chi}}\left(\frac{A^{2}\sigma_{\chi n}}{4\pi\mu_{\chi n}^{2}}\right)\int\mathrm{d}^{3}q\,\int\mathrm{d}^{3}w\,f_{\chi}\left(\mathbf{w};\,\hat{\mathbf{q}}\right)\,\delta\left(\frac{q^{2}}{2\mu_{\chi N}}+\mathbf{q}\cdot\mathbf{w}+\Delta E_{1\rightarrow2}\right)\left|\left\langle \psi_{2}\right|e^{i\mathbf{q}_{e}\cdot\hat{\mathbf{r}}}\left|\psi_{1}\right\rangle \right|^{2}\nonumber \\
\nonumber \\
 & = & \frac{\rho_{\chi}}{m_{\chi}}\left(\frac{A^{2}\sigma_{\chi n}}{4\pi\mu_{\chi n}^{2}}\right)\int\mathrm{d}^{3}q\,\int\frac{g_{\chi}\left(\mathbf{w};\,\hat{\mathbf{q}}\right)}{q\,w}\,\mathrm{d}w\,\mathrm{d}\phi_{\mathbf{\hat{\mathbf{q}}}\mathbf{w}}\,\Theta\left[w-w_{\mathrm{min}}\left(q,\,\Delta E_{1\rightarrow2}\right)\right]\left|\left\langle \psi_{2}\right|e^{i\mathbf{q}_{e}\cdot\hat{\mathbf{r}}}\left|\psi_{1}\right\rangle \right|^{2},
\label{eq:R_12}
\end{eqnarray}
where  $\rho_{\chi}$ and $m_{\chi}$ represent
the DM local density and the DM mass, respectively, $A$ is the atomic number of the target nucleus, $\sigma_{\chi n}$ is the DM-nucleon cross section, the bracket $\left\langle \cdots\right\rangle $ denotes the
average over the DM velocity distribution, $f_{\chi}\left(\mathbf{w},\,\hat{\mathbf{q}}\right)$
is the DM velocity distribution with unit vector $\hat{\mathbf{q}}$
as its zenith direction, $g_{\chi}\left(\mathbf{w};\,\hat{\mathbf{q}}\right)\equiv w^{2}f_{\chi}\left(\mathbf{w};\,\hat{\mathbf{q}}\right)$,
and $\Theta$ is the Heaviside step function. While $\phi_{\mathbf{\hat{\mathbf{q}}}\mathbf{w}}$
is the azimuthal angle of the spherical coordinate system $\left(\mathbf{\hat{\mathbf{q}}};\mathbf{w}\right)$,
the polar angle $\mathrm{d}\cos\theta_{\hat{\mathbf{q}}\mathbf{w}}$ has integrated
out the delta function in above derivation. $\mu_{\chi n}=m_{\chi}\,m_{n}/\left(m_{\chi}+m_{n}\right)$
($\mu_{\chi N}=m_{\chi}\,m_{N}/\left(m_{\chi}+m_{N}\right)$) is the
reduced mass of the DM-nucleon (DM-nucleus) pair, and $\Delta E_{1\rightarrow2}$
denotes the relevant energy difference. For the given $q$ and $\Delta E_{1\rightarrow2}$,
function $w_{\mathrm{min}}$ determines the minimum kinetically possible
velocity for the transition:
\begin{eqnarray}
w_{\mathrm{min}}\left(q,\,\Delta E_{1\rightarrow2}\right) & = & \frac{q}{2\,\mu_{\chi N}}+\frac{\Delta E_{1\rightarrow2}}{q}.
\end{eqnarray}
In practice, we take $\rho_{\chi}=0.3\,\mathrm{GeV/cm^{3}}$, and  the velocity distribution can be approximated as a truncated
Maxwellian form in the galactic rest frame, \textit{i.e.}, $f_{\chi}\left(\mathbf{w},\,\hat{\mathbf{q}}\right)\propto\exp\left[-\left|\mathbf{w}+\mathbf{v}_{\mathrm{e}}\right|^{2}/v_{0}^{2}\right]\,\Theta\left(v_{\mathrm{esc}}-\left|\mathbf{w}+\mathbf{v}_{\mathrm{e}}\right|\right)$,
with the earth's velocity $v_{\mathrm{e}}=230\,\mathrm{km/s}$, the
dispersion velocity $v_{0}=220\,\mathrm{km/s}$ and the galactic escape
velocity $v_{\mathrm{esc}}=544\,\,\mathrm{km/s}$.

\subsection{\label{sub:TB argument}Migdal effect with tight-binding approximation}

It is a natural idea to extend the above Migdal approach to the electronic
bands in the crystalline solids. However, due to the non-local nature
of the itinerant electrons in solids such extension does not seem
so straightforward, especially for the low energy excitation processes.
In order to apply the Migdal approach for the localized electron system
to the non-local electrons in  crystal, we manage to describe the
electrons with the Wannier functions (WFs) in the context of the tight-binding
(TB) approximation, in which the extensive nature of itinerant electrons
is encoded in the hopping integral.

Our strategy is outlined as follows. First, for simplicity it is assumed
that there is only one atom in each primitive unit cell, and
we express the Bloch wavefunction of an isolated electronic band $\left\{ \ket{i\,\mathbf{k}}\right\} $
(with band index $i$ and crystal momentum $\mathbf{k}$) in terms
of a complete set of localized WFs $\left\{ \ket{\mathbf{R}\,i}\right\} $
(with band index $i$ and cell index $\mathbf{\mathbf{R}}$) as
\begin{eqnarray}
\ket{i\,\mathbf{k}} & = & \sum_{\mathbf{R}}\frac{e^{i\mathbf{k}\cdot\mathbf{R}}}{\sqrt{N}}\ket{\mathbf{R}\,i},\label{eq:bloch}
\end{eqnarray}
where $N$ is the number of unit cells, or equivalently, the number
of mesh points in the first Brillouin zone (1BZ). The orthonormality of
WFs $\left\{ \ket{\mathbf{R}\,i}\right\} $, (\textit{i.e.}, $\ensuremath{\braket{\mathbf{R}'\,i'|\mathbf{R}\,i}}=\delta_{i'i}\,\delta_{\mathbf{R}'\mathbf{R}}$)
corresponds to the normalization convention over the whole crystal
such that $\ensuremath{\braket{i'\,\mathbf{k}'|i\,\mathbf{k}}}=\delta_{i'i}\,\delta_{\mathbf{k}'\mathbf{k}}$.
Accordingly it is straightforward to obtain the inverse relation
\begin{eqnarray}
\ket{\mathbf{R}\,i} & = & \sum_{\mathbf{\mathbf{k}}}\frac{e^{-i\mathbf{k}\cdot\mathbf{R}}}{\sqrt{N}}\ket{i\,\mathbf{k}}.\label{eq:WannierFunction0}
\end{eqnarray}
Next, we impose the Galilean boost operator $\hat{G}_{\mathbf{R}}\left(\mathbf{q}_{e}\right)\equiv e^{i\mathbf{q}_{e}\cdot\hat{\mathbf{r}}}$
exclusively on the recoiled atom located at $\mathbf{R}$, with velocity
$-\mathbf{q}/m_{N}$. However, such extensive use of operator $e^{i\mathbf{q}_{e}\cdot\hat{\mathbf{r}}}$
in crystalline environment needs to be carefully examined, considering that nuclei couple with each other in the crystal structure, and
the recoiled nucleus no longer amounts to an evident reference to describe
the electronic excitation process while ambient nuclei remain at rest.
To this point we make some detailed explanation. First,  the effectiveness of the impulse approximation where the recoiled nucleus is treated as free particle requires that the time scale of the hard scattering is much smaller than that of the atomic vibrations, or equivalently, that the energy deposition is much higher than the typical phonon energy~\cite{Trickle:2019nya}, which translates into a transferred momentum  $q\gg\sqrt{2\,\omega_{\mathrm{D}}\,m_{N}}\sim\mathcal{O}\left(10\,\mathrm{keV}\right)$, with the Debye frequency of the system $\omega_{\mathrm{D}}\sim\mathcal{O}\left(10^{-1}\,\mathrm{eV}\right)$. On the other hand, a momentum
transfer larger than $q\sim\mathcal{O}\left(\mathrm{keV}\right)$ is sufficient to resolve
the diamond structure, and in this case the Galilean transformation
can be imposed onto a specific atom. Then imagine the atom residing
at $\mathbf{R}=\mathbf{0}$ is struck, from Eq.~(\ref{eq:bloch})
we assume that the instantaneous eigenstate responding to this recoiled
atom takes the form
\begin{eqnarray}
\left|i\,\mathbf{k}\right\rangle _{\mathrm{struck}} & = & \frac{1}{\sqrt{N}}\left(\hat{G}_{\mathbf{0}}\left(-\mathbf{q}_{e}\right)\left|\mathbf{0}\,i\right\rangle +\sum_{\mathbf{R}\neq\mathbf{0}}e^{i\mathbf{k}\cdot\mathbf{R}}\ket{\mathbf{R}\,i}\right),
\end{eqnarray}
while the actual electronic state after the collision remains intact
(\textit{i.e.}, $\ket{i\,\mathbf{k}}$) under the sudden approximation.
Secondly, since the Debye frequency of the system
is much lower than the energy gap $ E_{g}\sim\mathcal{O}\left(\mathrm{\mathrm{eV}}\right)$ between
valence and conduction bands, the evolution of the eigenstates divided by the band gap
can be approximated as adiabatic. In other words, there will be no transition from the valence bands to the conduction bands during the evolution of the Hamiltonian because the vibrational frequency of the perturbed nucleus is too low to excite an electron across the band gap. As the recoiled nucleus dissipates its energy in the form of phonons, these eigenstates eventually evolve adiabatically back to the original ones. So  transitions are exclusively attributed to the sudden boost of the recoiled nucleus.  In this sense, the state $\ket{i\,\mathbf{k}}$
is regarded as perturbed with respect to the adiabatically evolving
Hamiltonian and can be projected to eigenstates $\left|i'\,\mathbf{k'}\right\rangle _{\mathrm{struck}}$
to derive the transition amplitude $_{\mathrm{struck}}\left\langle i'\,\mathbf{k}'\right|\left.i\,\mathbf{k}\right\rangle $. Thus, as a direct extension of the Migdal effect
in atoms, the transition amplitude between an initial valence state
and final conducting state  $\left(i'\neq i\right)$ can be written as\footnote[3]{\renewcommand{\baselinestretch}{1}\selectfont In derivation one just adds and subtracts an identity operator, and exploits the facts that the initial and final states are orthogonal,  and that  the Galilean operator reduces to a unit operator for $\ket{\mathbf{R}'\,i'}$~($\mathbf{R}'\neq\mathbf{0}$).}
\begin{eqnarray}
_{\mathrm{struck}}\left\langle i'\,\mathbf{k}'\right|\left.i\,\mathbf{k}\right\rangle  & = & \left\langle i\,\mathbf{k}\right|\hat{G}_{\mathbf{0}}\left(-\mathbf{q}_{e}\right)\left|i'\,\mathbf{k}'\right\rangle ^{*} \nonumber \\
 & = & \frac{1}{N}\sum_{\mathbf{R}}\left\langle \mathbf{R}\,i\right|\hat{G}_{\mathbf{0}}\left(-\mathbf{q}_{e}\right)-\mathbf{1}\left|\mathbf{0}\,i'\right\rangle ^{*}e^{i\mathbf{k}\cdot\mathbf{R}}
\nonumber \\
 & = & \frac{1}{N}\sum_{\mathbf{R}}\left\langle \mathbf{0}\,i'\right|e^{i\mathbf{q}_{e}\cdot\hat{\mathbf{r}}}-\mathbf{1}\left|\mathbf{R}\,i\right\rangle e^{i\mathbf{k}\cdot\mathbf{R}}
 ,\label{eq:transitionAmplitude}
\end{eqnarray}
where the hopping integrals between neighboring atoms $\left\langle \mathbf{0}\,i'\right|e^{i\mathbf{q}_{e}\cdot\hat{\mathbf{r}}}\left|\mathbf{R}\,i\right\rangle $
reflect the delocalized nature of the electrons in crystalline solids.
\begin{figure}
\begin{centering}
\includegraphics[scale=0.4]{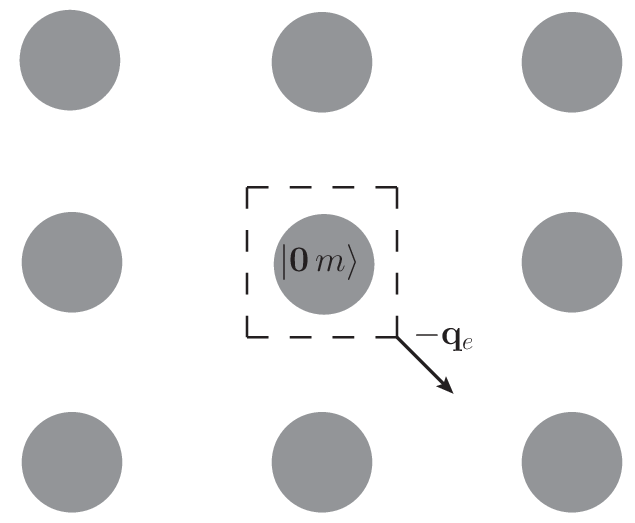}\hspace{3cm}\includegraphics[scale=0.75]{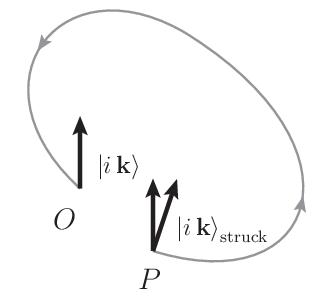}
\par\end{centering}

\caption{\textit{Left}: Schematic illustration of our proposal to describe
the electronic excitation induced by the struck nucleus located
at $\mathbf{R}=\mathbf{0}$: only the initial states $\left\{ \ket{\mathbf{0}\,m}\right\} $
are relevant for the excitation process and thus are subjected to the
Galilean transformation $\hat{G}_{\mathbf{0}}\left(-\mathbf{q}_{e}\right)$.
\textit{Right}: Conceptual illustration of the evolution of Hamiltonian
in parameter space (\textit{gray} \textit{curve}), with point $O$
($P$) corresponding to the original (perturbed) Hamiltonian and relevant
eigenstates $\left\{ \ket{i\,\mathbf{k}}\right\} $ ($\left\{ \left|i\,\mathbf{k}\right\rangle _{\mathrm{struck}}\right\} $).
These two points are equivalent in the sense that they are connected
to each other through an adiabatic evolution. See text for details.}
\end{figure}
 In addition,
since the timescale of the excitation is roughly $E_{g}^{-1}$, during
which the recoiled nucleus at most travels a distance around $\left(q/m_{N}\right)E_{g}^{-1}\sim\mathcal{O}\left(10^{-2}\,\textrm{\AA}\right)$
for a momentum transfer $q\sim1\,\mathrm{MeV}$,  the excitation
can be regarded as instantaneous, and thus the configuration effect of
the displaced  nucleus can be ignored.

\section{\label{sec:WFs in Migdal Effects}Practical calculation of excitation
event rates with WFs}

\subsection{Formalism}

In this section we will derive the formalism of the excitation event
rate induced by a recoiled nucleus in the bulk  diamond. Here we consider
a more realistic multiband case where a separate group of $J$ bands
cross with each other, and a Bloch orbital can be expressed in terms
of WFs $\left\{ \ket{\mathbf{R}\,m}\right\} $ in following way:
\begin{eqnarray}
\ket{i\,\mathbf{k}} & = & \sum_{\mathbf{R}}\frac{e^{i\mathbf{k}\cdot\mathbf{R}}}{\sqrt{N}}\left(\sum_{m=1}^{J}U_{mi}^{\left(\mathbf{k}\right)\dagger}\ket{\mathbf{R}\,m}\right),\label{eq:Bloch WF}
\end{eqnarray}
and its inverse transformation
\begin{eqnarray}
\ket{\mathbf{R}\,m} & = & \sum_{\mathbf{\mathbf{k}}}\frac{e^{-i\mathbf{k}\cdot\mathbf{R}}}{\sqrt{N}}\left(\sum_{i=1}^{J}U_{im}^{\left(\mathbf{k}\right)}\ket{i\,\mathbf{k}}\right),\label{eq:WannierFunction}
\end{eqnarray}
with the unitary matrix $\mathbf{U}^{\left(\mathbf{k}\right)}$ that
mixes different bands for each $\mathbf{k}$-vector. In practice,
we use the code $\mathtt{Wannier\,90}$~\cite{MOSTOFI20142309} to
realize this scheme, which specializes in constructing the maximally
localized Wannier functions (MLWFs) from a set of Bloch states. To
avoid distraction from present discussion, we arrange a short review of
the MLWF in the Appendix~\ref{sec:Appendixa}. In order to transplant
the treatment on the Migdal effect applied in the isolated atom to
the crystalline environment, the WFs are required to be atom-centered.
Since there are two atoms in the unit cell of the crystalline diamond, here
we introduce the Galilean boost operator $\hat{G}_{\mathbf{0}1}\left(\mathbf{q}_{e}\right)$
that accounts for recoil effect of the first diamond atom at the site
$\left(\mathbf{R}=\mathbf{0},\,\boldsymbol{\tau}_{1}=\mathbf{0}\right)$,
so one has the following transition amplitude $\left(i'\neq i\right)$:
\begin{eqnarray}
\left\langle i\,\mathbf{k}\right|\hat{G}_{\mathbf{0}1}\left(-\mathbf{q}_{e}\right)\left|i'\,\mathbf{k}'\right\rangle ^{*}  & = & \frac{1}{N}\sum_{m,m'}\sum_{\mathbf{R}}U_{mi}^{\left(\mathbf{k}\right)\dagger}\left\langle \mathbf{R}\,m\right|\hat{G}_{\mathbf{0}1}\left(-\mathbf{q}_{e}\right)-\mathbf{1}\left|\mathbf{0}\,m'\right\rangle ^{*}\left(U_{m'i'}^{\left(\mathbf{k}'\right)\dagger}\right)^{*}e^{i\mathbf{k}\cdot\mathbf{R}}\nonumber \\
\nonumber \\
 & \simeq & \frac{i}{N}\mathbf{q}_{e}\cdot\left(\sum_{m_{1},m}\sum_{\mathbf{R}}U_{i'\,m_{1}}^{\left(\mathbf{k}'\right)}\left\langle \mathbf{0}\,m_{1}\right|\hat{\mathbf{r}}\left|\mathbf{R}\,m\right\rangle U_{mi}^{\left(\mathbf{k}\right)\dagger}e^{i\mathbf{k}\cdot\mathbf{R}}\right)\nonumber \\
 & = & \frac{i}{N}\mathbf{q}_{e}\cdot\mathbf{J}_{1\left(i'\mathbf{k}';i\mathbf{k}\right)},\label{eq:r_matrix}
\end{eqnarray}
where the origin of coordinate operator $\hat{\mathbf{r}}$ is placed
at the atom 1 in the unit cell, and the partial summation
 over  $m_{1}$ corresponds to the WFs centered at atom 1. We use $\mathbf{J}_{1\left(i'\mathbf{k}';i\mathbf{k}\right)}$
to denote the summation in parenthesis in the second line.  Since momentum transfer $\mathbf{q}$ is highly suppressed by $m_{e}/m_{N}$, in derivation we assume $\left|\mathbf{q}_{e}\right|\cdot\left|\mathbf{r}\right|\ll1$, which is a good approximation for a  sub-GeV DM. Similar discussion
can be easily applied to the second atom located at $\boldsymbol{\tau}_{2}=\left(1/4,\,1/4,\,1/4\right)$,
only keep in mind that the origin of operator $\hat{\mathbf{r}}$
is also placed at atom 1 in practical use of $\mathtt{Wannier\,90}$.
After a translation, the corresponding transition amplitude for the
atom 2 at the site $\mathbf{R}=\mathbf{0}$ is modified as
\begin{eqnarray}
\left\langle i\,\mathbf{k}\right|\hat{G}_{\mathbf{0}2}\left(-\mathbf{q}_{e}\right)\left|i'\,\mathbf{k}'\right\rangle ^{*}  & \simeq & \frac{i}{N}\mathbf{q}_{e}\cdot\left(\sum_{m_{2},m}\sum_{\mathbf{R}}U_{i'\,m_{2}}^{\left(\mathbf{k}'\right)}\left\langle \mathbf{0}\,m_{2}\right|\hat{\mathbf{r}}\left|\mathbf{R}\,m\right\rangle U_{mi}^{\left(\mathbf{k}\right)\dagger}e^{i\mathbf{k}\cdot\mathbf{R}}-
\mathbf{d}\sum_{m_{2}}U_{i'\,m_{2}}^{\left(\mathbf{k}'\right)}U_{m_{2}i}^{\left(\mathbf{k}\right)\dagger}\right)\nonumber \\
\nonumber \\
 & = & \frac{i}{N}\mathbf{q}_{e}\cdot\mathbf{J}_{2\left(i'\mathbf{k}';i\mathbf{k}\right)},
\end{eqnarray}
with $\mathbf{d}$ being the position vector of atom 2 relative to
atom 1, and $\mathbf{J}_{2\left(i'\mathbf{k}';i\mathbf{k}\right)}$
encodes the terms in parenthesis. Thus, after taking into account
the two degenerate spin states, the total excitation event rate can
be expressed as
\begin{eqnarray}
R & = & \frac{\rho_{\chi}}{m_{\chi}}\left(\frac{A^{2}\sigma_{\chi n}}{2\pi\mu_{\chi n}^{2}}\right)\left(\frac{m_{e}}{m_{N}}\right)^{2}\frac{V^{2}}{N}\int\mathrm{d}^{3}q\,\sum_{i'}^{c}\sum_{i}^{v}\int_{1\mathrm{BZ}}\frac{\mathrm{d}^{3}k'}{\left(2\pi\right)^{3}}\frac{\mathrm{d}^{3}k}{\left(2\pi\right)^{3}}\,\left\{ \int\frac{g_{\chi}\left(\mathbf{w},\,\hat{\mathbf{q}}\right)}{q\,w}\,\mathrm{d}w\,\mathrm{d}\phi_{\mathbf{\hat{\mathbf{q}}}\mathbf{w}}\,\right.\nonumber \\
 &  & \times\Theta\left[w-w_{\mathrm{min}}\left(q,\,E_{i'\mathbf{k}'}-E_{i\mathbf{k}}\right)\right]\,\left.\left(\left|\mathbf{q}\cdot\mathbf{J}_{1\left(i'\mathbf{k}';i\mathbf{k}\right)}\right|^{2}+\left|\mathbf{q}\cdot\mathbf{J}_{2\left(i'\mathbf{k}';i\mathbf{k}\right)}\right|^{2}\right)\right\} ,
\label{eq:EventRate0}
\end{eqnarray}
where the sums are over the valence bands for initial states and the
conducting bands for final states, respectively. For simplicity we approximate the velocity distribution
as an isotropic one, and as a result the angular correlation between the laboratory velocity  with respect to the galaxy and the orientation of the crystal is eliminated.  Besides, in order to make the scan of parameters computationally more efficient, Eq.~(\ref{eq:EventRate0}) can be further equivalently expressed as
\begin{eqnarray}
R & = & \frac{\rho_{\chi}}{m_{\chi}}\left(\frac{A^{2}\sigma_{\chi n}\,q_{\mathrm{ref}}^{2}}{3\,\mu_{\chi n}^{2}}\right)\left(\frac{m_{e}}{m_{N}}\right)^{2}N\int\frac{4\pi g_{\chi}\left(w\right)}{w}\,\,\mathrm{d}w\int\mathrm{d}\ln E_{e}\,\mathrm{d}\ln q\,\,\Theta\left[w-w_{\mathrm{min}}\left(q,\,E_{e}\right)\right]\,\mathcal{F}\left(q,\,E_{e}\right),
\end{eqnarray}
where a momentum reference value is constructed as $q_{\mathrm{ref}}=2\pi/a$,
and a non-dimensional crystal form factor is introduced as
\begin{eqnarray}
\mathcal{F}\left(q,\,E_{e}\right) & \equiv & \left(\frac{q}{q_{\mathrm{ref}}}\right)^{2}\sum_{i'}^{c}\sum_{i}^{v}\int_{1\mathrm{BZ}}\frac{\varOmega\,\mathrm{d}^{3}k'}{\left(2\pi\right)^{3}}\frac{\varOmega\,\mathrm{d}^{3}k}{\left(2\pi\right)^{3}}\,\left[E_{e}\,\delta\left(E_{i'\mathbf{k}'}-E_{i\mathbf{k}}-E_{e}\right)\,q^{2}\left(\left|\mathbf{J}_{1\left(i'\mathbf{k}';i\mathbf{k}\right)}\right|^{2}+\left|\mathbf{J}_{2\left(i'\mathbf{k}';i\mathbf{k}\right)}\right|^{2}\right)\right].
\label{eq:crystal Form Factor}
\end{eqnarray}
Specifically, after inserting the monochromatic velocity distribution $4\pi g_{\chi}\left(v\right)=\delta\left(v-w\right)$
the total transition rate for parameter pair $(m_{\chi},\,w)$ can
be recast as
\begin{eqnarray}
R\left(m_{\chi},\,w\right) & = & \frac{\rho_{\chi}}{m_{\chi}}\left(\frac{A^{2}\sigma_{\chi n}\,q_{\mathrm{ref}}^{2}}{3\,w\,\mu_{\chi n}^{2}}\right)\left(\frac{m_{e}}{m_{N}}\right)^{2}N\int\mathrm{d}\ln E_{e}\,\mathrm{d}\ln q\,\Theta\left[w-w_{\mathrm{min}}\left(q,\,E_{e}\right)\right]\,\mathcal{F}\left(q,\,E_{e}\right).\label{eq:EvenRateScan}
\end{eqnarray}

\subsection{\label{sub:ComputationalDetails}Computational details and results}

\begin{figure}
\begin{centering}
\includegraphics[scale=0.55]{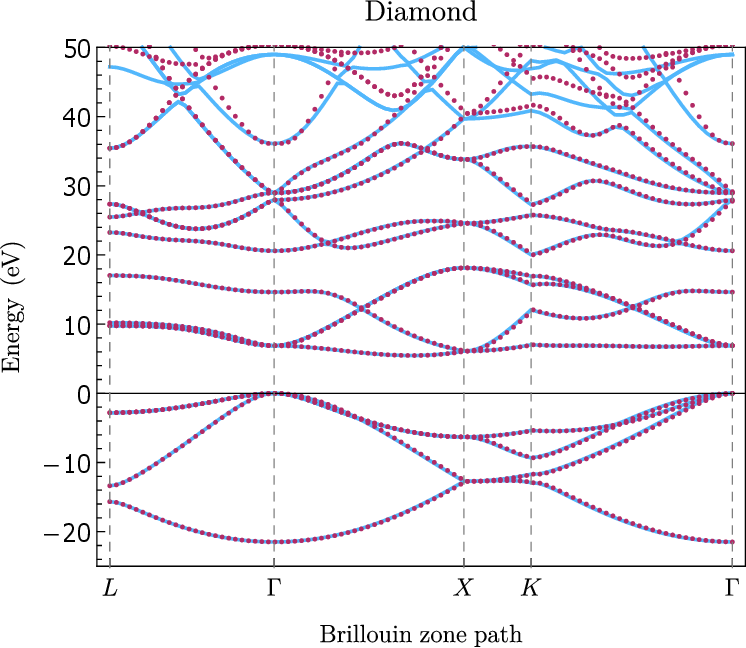}\hspace{1cm}\includegraphics[scale=0.195]{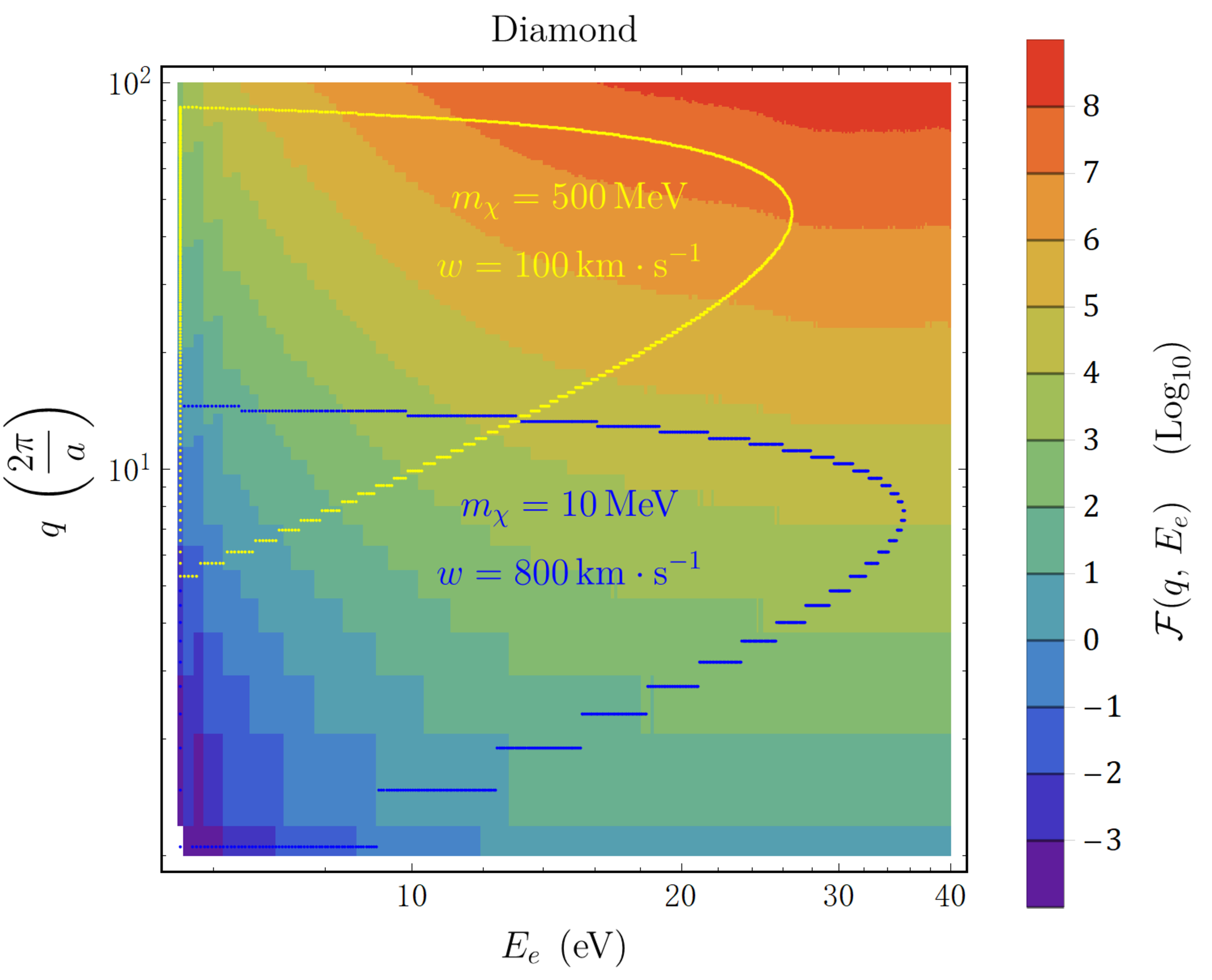}
\par\end{centering}

\caption{\textit{\label{fig:Band}Left: }The band structures of bulk crystalline
diamond obtained from DFT calculation (blue solid line) and Wannier
interpolation (red dotted line). \textit{Right}: The crystal form
function $\mathcal{F}\left(q,\,E_{e}\right)$. The area contoured
in \textit{yellow} (\textit{blue}) represents the parameter region
relevant for the Migdal excitation event rate for parameter $m_{\chi}=500\,\mathrm{MeV}$,
$w=100\,\mathrm{km/s}$ ($m_{\chi}=10\,\mathrm{MeV}$, $w=800\,\mathrm{km/s}$).
See text for details.}
\end{figure}

Now we put into practice the estimate of the Migdal excitation
event rate. With $\mathtt{Quantum\,Espresso}$ code~\cite{Giannozzi_2009},
we first perform the DFT calculation to obtain the Bloch eigenfunctions
and eigenvalues using the plane-wave basis set and Troullier-Martins
norm-conserving (NC) pseudopotentials in the Kleinman-Bylander representation,
on a 20$\times$20$\times$20 Monkhorst-Pack mesh of $k$-points.
The exchange-correlation functional is treated within the generalized
gradient approximation (GGA) parametrized by Perdew, Burke, and Ernzerhof
(PBE)~\cite{PhysRevLett.77.3865}. The energy cut is set to $90$ Ry and a lattice constant $a=3.560$~\AA~
obtained from relaxation is adopted.

Then we invoke the software package $\mathtt{Wannier\,90}$~\cite{MOSTOFI20142309}
to compute the matrix element $\left\langle \mathbf{0}\,m'\right|\hat{\mathbf{r}}\left|\mathbf{R}\,m\right\rangle $
and the unitary matrix $\mathbf{U}^{\left(\mathbf{k}\right)}$ using a smaller homogeneous set of 16$\times$16$\times$16 $k$-points. We
generate $J=32$ WFs out of $\mathcal{J}_{\mathbf{k}}=72$ Bloch wavefunctions
from the DFT calculation by beginning with a set of 32 localized
trial orbitals $\left\{ g_{n}\left(\mathbf{r}\right)\right\} $ that
correspond to $s$, $p$, $d$, $f$ orbitals as some rough initial
guess for these WFs\footnote[4]{\renewcommand{\baselinestretch}{1}\selectfont See Appendix~\ref{sec:Appendixa} for a brief review on the  MLWFs.}. In order to fix the WFs at the atom while containing
their spreads, the gauge selection step is expediently spared because
otherwise some MLWFs will be found located at interstitial sites,
which makes the picture of recoiled atom ambiguous. The widest spread of these generated WFs is around $\left(1.4\,\textrm{\AA}\right)^{2}$, so hopping terms
within up to the third neighbor unit cells
are  sufficient to guarantee convergence in calculation of $\left\langle \mathbf{0}\,m'\right|\hat{\mathbf{r}}\left|\mathbf{R}\,m\right\rangle $.

On the other hand, once a set of localized WFs have been determined, and hence the Hamiltonian matrices $\left\langle \mathbf{R}\,m\right|\hat{H}\left|\mathbf{0}\,m'\right\rangle $
are tabulated, the band structures become an eigenvalue problem

\begin{eqnarray}
\sum_{m'=1}^{J}\sum_{\mathbf{R}}\left\langle \mathbf{R}\,m\right|\hat{H}\left|\mathbf{0}\,m'\right\rangle e^{-i\mathbf{k}\cdot\mathbf{R}}\,U_{m'i}^{\left(\mathbf{k}\right)\dagger} & = & \bar{\epsilon}_{i\,\mathbf{k}}\,U_{mi}^{\left(\mathbf{k}\right)\dagger},
\end{eqnarray}
where the eigenvalue $\bar{\epsilon}_{i\,\mathbf{k}}$ corresponds
to the $i$-th band energy at $\mathbf{k}$, and the eigenvectors
$U_{mi}^{\left(\mathbf{k}\right)\dagger}$ that form a $J\times J$
unitary matrix are deliberately written in accordance with Eq.~(\ref{eq:Bloch WF}).
Such procedure is called the Wannier interpolation. The band structures of crystalline diamond from the DFT calculation
and interpolation are presented in the left panel of Fig.~\ref{fig:Band},
where the blue solid line (red dotted line) represents the DFT calculation
(Wannier interpolation). To reproduce exactly the original DFT band
structures in a specific energy range, Bloch states spanning relevant
range are faithfully retained in the subspace selection procedure, and
such energy range is referred to as ``frozen energy window'', or
``inner window''. In this work, we choose a frozen energy window
ranging from the bottom of the valence band to 40 eV above the valence
band maximums, which is also reflected in Fig.~\ref{fig:Band}.

In practical computation of the form factor Eq.~(\ref{eq:crystal Form Factor}) we use a  bin width
$\Delta E=0.059\,\mathrm{eV}$  to smear the delta function,
and the integrand is evaluated at the central value in each energy
bin. Besides, the integrals of the continuous $k$-points in the 1BZ are replaced by corresponding summations over the uniform 16$\times$16$\times$16 $k$-points. In the right panel of Fig.~\ref{fig:Band} shown is the crystal form
factor $\mathcal{F}\left(q,\,E_{e}\right)$ introduced in Eq.~(\ref{eq:crystal Form Factor}),
where momentum transfer $q$ is expressed in terms of the diamond
reciprocal lattice $2\pi/a\approx3.48\,\mathrm{keV}$. For demonstration,
we choose two benchmark pairs of parameters $\left(m_{\chi},\,w\right)$
to calculate the transition event rate in Eq.~(\ref{eq:EvenRateScan}).
Due to the step function that embodies the kinetic constraint, only
the areas enclosed in the contours contribute to the excitation event
rate. For example, when a cross section $\sigma_{\chi n}=10^{-38}\,\mathrm{cm}^{2}$
is assumed, parameters $m_{\chi}=500\,\mathrm{MeV},\:w=100\,\mathrm{km/s}$
presented in yellow corresponds to an excitation event rate of $0.027\,/\mathrm{kg/yr}$,
while $m_{\chi}=10\,\mathrm{MeV},\:w=800\,\mathrm{km/s}$ in blue
corresponds to an excitation event rate of $0.299\,\mathrm{/kg/yr}$.
It is noted that although a heavier DM tends to prompt a larger momentum
transfer and hence results in a larger transition probability, a suppression
factor inversely proportional to $m_{\chi}^{3}$ may alleviate or
even offset such mass effect in the sub-GeV mass range, depending
on velocity.

\begin{figure}
\begin{centering}
\includegraphics[scale=0.7]{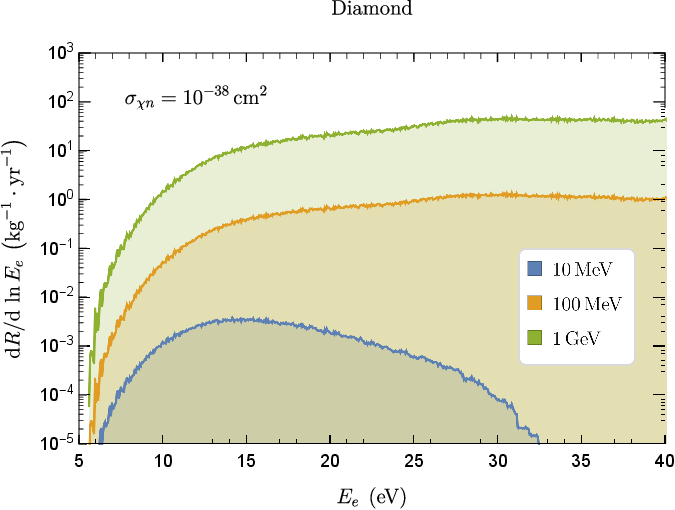}\hspace{0.5cm}\includegraphics[scale=0.72]{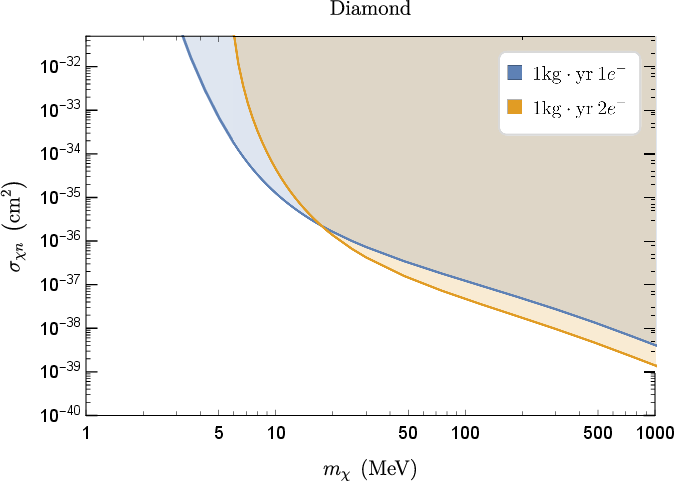}
\par\end{centering}

\caption{\label{fig:Energy Spectrum}\textit{Left}: The differential electronic
excitation event rate of the Migdal effect in crystalline
diamond for reference values $\sigma_{\chi n}=10^{-38}\,\mathrm{cm}^{2}$
and $m_{\chi}=10\,\mathrm{MeV}$ (\textit{blue}), $100\,\mathrm{MeV}$(\textit{orange})
and $1\,\mathrm{GeV}$ (\textit{green}), respectively. \textit{Right}: Cross-section
sensitivities for the Migdal effect at 90\% C.L. for a 1 kg-yr diamond
detector, based on a single-electron (\textit{blue}) and a two-electron
(\textit{orange}) signals, respectively. See text for details.}
\end{figure}
In left panel of Fig.~\ref{fig:Energy Spectrum} we plot the velocity-averaged
energy spectra of the Migdal excitation for DM mass $m_{\chi}=10\,\mathrm{MeV}$
(\textit{blue}), $100\,\mathrm{MeV}$(\textit{orange}) and $1\,\mathrm{GeV}$
(\textit{green}), respectively. It is observed that these spectra
do not fall off as quickly as the case in direct DM-electron excitation
process calculated in Ref.~\cite{Essig:2015cda}. Hence a wider energy range is usually required to fully describe
the Migdal effect in crystalline environment. From the energy spectra
we also estimate the sensitivity of  1 kg-yr exposure of diamond detector in
the right panel of Fig.~\ref{fig:Energy Spectrum}, assuming an average
energy of $13\,\mathrm{eV}$ for producing one electron-hole pair~\cite{Kurinsky:2019pgb}.
The 90\% C.L. exclusion contours for  DM-nucleon cross section for both the
single-electron (\textit{blue}) and the two-electron (\textit{orange})
bins are presented with   no background event assumed.

\section{\label{sec:Conclusions}Summary and discussions}

In this paper we presented a tight-binding approach to describe the
Migdal effect in the diamond crystal. This localized description is
a natural choice to generalize the well established treatment for
isolated atoms to the case in crystalline solids. To achieve such TB description
we generate a set of atom-centered WFs by use of software packages
$\mathtt{Quantum\,Espresso}$ and $\mathtt{Wannier\,90}$. While the
localization effect of the recoiled atom is preserved in forms of
WFs, the delocalized nature of the Bloch states in solids is encoded
in the hopping terms at the same time. Based on these hopping integrals,
the electronic excitation rates induced by the recoiled ion were computed
straightforwardly.

Here we make some comments on our methodology. As has been noted in
previous section, our method deviates a little from the standard procedure
to derive the MLWFs implemented in $\mathtt{Wannier\,90}$, which
usually includes two steps, namely, subspace selection and gauge selection\footnote[5]{\renewcommand{\baselinestretch}{1}\selectfont See Appendix~\ref{sec:Appendixa} for more details.}.
The latter step is skipped in our computation so as to restrict the
centers of WFs to the lattice sites. As a consequence, the spreads
of those WFs have not been optimally minimized. From a conceptual
point of view, this does not cause a severe problem because although
WFs indicate some atomic properties and provide intuitive pictures
of the chemical bonds, they do not necessarily bear definite physical
meanings such as real atomic wavefunctions in solids. Actually, to
determine the WFs is experimentally infeasible, even in principle~\cite{RevModPhys.84.1419}.
In this sense, even a set of relatively ``fat'' WFs suffice in calculation
of the excitation event rate, as long as sufficiently large number
of neighbors are  included in hopping integrals. However, WFs with
too large spreads indeed cause inconvenience in practice. This partly
explains why we choose diamond crystal rather than other tetrahedral semiconductors
as example to describe the Migdal effect in semiconductors:
for the cases of silicon and germanium, $\mathtt{Wannier\,90}$ is found to
generate atom-centered WFs with much larger spreads in the same energy window unless a gauge optimization is performed,
and hence more computational efforts are required to reach convergence.

Another concern is how the calculation of Migdal excitation rates
will rely on the selection of the WFs, considering there existing
a gauge freedom in the unitary matrix $\mathbf{U}^{\left(\mathbf{k}\right)}$ in Eq.~(\ref{eq:WannierFunction}). In fact, since it is required that the centers of
the WFs locate at the lattice sites to account for the struck atoms,
such invariance of arbitrary WF selection is unnecessary in formulating
the transition probability. To see this, recall the transition amplitude
in Eq.~(\ref{eq:r_matrix}), while the initial  state $ \sum\limits _{m,\mathbf{R}}e^{i\mathbf{k}\cdot\mathbf{R}}\left|\mathbf{R}\,m\right\rangle U_{mi}^{\left(\mathbf{k}\right)\dagger}\propto\left|i\,\mathbf{k}\right\rangle $
as a whole is independent of the WFs, the conduction final part $\sum\limits _{m_{1}}U_{i'\,m_{1}}^{\left(\mathbf{k}'\right)}\left\langle \mathbf{0}\,m_{1}\right| $
is invariant only within the WF subspace attached to atom 1. This
can be understood intuitively with the example of tetrahedral $sp^{3}$
hybridization, where the valence wavefunctions around an atom can
be expressed either as linear combinations of $s$ and $p$ atomic
orbitals, or combinations of four identical $sp$-hybrid orbitals
pointing along the directions from the center to the corners of a
tetrahedron. All these different choices should keep the sum $\sum\limits _{m_{1}}U_{i'\,m_{1}}^{\left(\mathbf{k}'\right)}\left\langle \mathbf{0}\,m_{1}\right| $
invariant. To verify this, we used $s$, $p$ atomic orbitals and
$sp^{3}$ hybrids among other orbitals as trial wavefunctions $\left\{ g_{n}\left(\mathbf{r}\right)\right\} $
 to wannierize and calculate the transition
event rate respectively, and the difference between the two results turns
out to be well within a few percent, indicating a good agreement.
Thus our approach has  proved both effective and self-consistent.

\appendix

\begin{acknowledgments}
This work was supported by Science Challenge Project under No. TZ2016001, by the National Key R\&D Program of China under Grant under No. 2017YFB0701502, and  by National Natural Science Foundation of  China under No. 11625415.
\end{acknowledgments}

\section{\label{sec:Appendixa}Maximally localized Wannier functions}
\setcounter{equation}{0}
\setcounter{subsection}{0}
\renewcommand{\theequation}{A.\arabic{equation}}
\renewcommand{\thesubsection}{A.\arabic{subsection}}

Since our practical realization of the atom-centered WFs is closely related to
the generation of the maximally localized Wannier functions~ (MLWFs) via implementing
software $\mathtt{Wannier\,90}$~\cite{MOSTOFI20142309}, here we
take a brief review. From Eq.~(\ref{eq:WannierFunction}) it is straightforward
to see that the WFs are non-unique due to the arbitrariness of the
unitary matrix $\mathbf{U}^{\left(\mathbf{k}\right)}$. In order to
overcome such indeterminacy, Mazari and Vanderbilt~\cite{PhysRevB.56.12847}
developed a procedure to minimize the second moment of the WFs around
their centers so as to generate a set of well-defined and localized
WFs, namely,  the maximally localized Wannier functions.
Given an isolated group of $J$ Bloch bands, the procedure begins
with a set of $J$ localized trial orbitals $\left\{ g_{n}\left(\mathbf{r}\right)\right\} $
as some rough initial guess for corresponding WFs, which are then
projected onto those Bloch wavefunctions as the following,
\begin{eqnarray}
\left|\phi_{n\mathbf{k}}\right\rangle  & = & \sum_{i=1}^{J}\left|i\,\mathbf{k}\right\rangle \braket{i\,\mathbf{k}|g_{n}},
\end{eqnarray}
which are typically smooth in $\mathbf{k}$ and hence are well-localized
in space. By further Löwdin-orthonormalizing these functions one obtains
$J$ Bloch-like states $\widetilde{\left|i\,\mathbf{k}\right\rangle }$
that are related to the original $\left|i\,\mathbf{k}\right\rangle $
via a $J\times J$ unitary matrix $U_{\mathbf{k}}$ in the following
manner,
\begin{eqnarray}
\widetilde{\left|i\,\mathbf{k}\right\rangle } & = & \sum_{j=1}^{J}\left|j\,\mathbf{k}\right\rangle U_{\mathbf{k},ij}.\label{eq:Bloch-like orbital}
\end{eqnarray}
Substituting $\left|i\,\mathbf{k}\right\rangle $ in Eq.~(\ref{eq:WannierFunction0})
with $\widetilde{\left|i\,\mathbf{k}\right\rangle }$ we then have
$J$ localized WFs. The minimization criterion proposed by Mazari
and Vanderbilt mentioned above is to minimize the functional
\begin{eqnarray}
\Omega & = & \sum_{n}\left[\left\langle \mathbf{0}\,n\right|\hat{r}^{2}\left|\mathbf{0}\,n\right\rangle -\left|\left\langle \mathbf{0}\,n\right|\hat{\mathbf{r}}\left|\mathbf{0}\,n\right\rangle \right|^{2}\right]\nonumber \\
 & = & \varOmega_{\mathrm{I}}+\tilde{\varOmega},
\end{eqnarray}
with
\begin{eqnarray}
\varOmega_{\mathrm{I}} & = & \sum_{n}\left[\left\langle \mathbf{0}\,n\right|\hat{r}^{2}\left|\mathbf{0}\,n\right\rangle -\sum_{\mathbf{R},m}\left|\left\langle \mathbf{R}\,m\right|\hat{\mathbf{r}}\left|\mathbf{0}\,n\right\rangle \right|^{2}\right]
\end{eqnarray}
and
\begin{eqnarray}
\tilde{\varOmega} & = & \sum_{n}\,\sum_{\mathbf{R}m\neq\mathbf{0}n}\left|\left\langle \mathbf{R}\,m\right|\hat{\mathbf{r}}\left|\mathbf{0}\,n\right\rangle \right|^{2}.
\end{eqnarray}
The purpose of such separation is that $\varOmega_{\mathrm{I}}$ is
invariant under arbitrary unitary transformation $\mathbf{U}^{\left(\mathbf{k}\right)}$
in Eq.~(\ref{eq:WannierFunction}), and hence the minimization of
$\Omega$ depends only on the variation effects of gauge $\mathbf{U}^{\left(\mathbf{k}\right)}$
on the term $\tilde{\varOmega}$. Since the matrix $U_{\mathbf{k}}$
in Eq.~(\ref{eq:Bloch-like orbital}) has already provided $J$ sufficiently
localized WFs, we use them as the starting point for the iterative
steepest-descent method to reach the optimal unitary transformation $\mathbf{U}^{\left(\mathbf{k}\right)}$
that minimizes $\tilde{\varOmega}$. This procedure is called gauge
selection.

In more general cases, additional modifications are required to pick
the Bloch orbitals of interest from those unwanted ones when the bands
are crossing with each other.  The Souza-Marzari-Vanderbilt
method proposed in Ref.~\cite{PhysRevB.65.035109} successfully
realized such disentanglement. In this approach, one first identifies
a set of $\mathcal{J}_{\mathbf{k}}\geq J$ Bloch orbitals that form
a $\mathcal{J}_{\mathbf{k}}$-dimensional Hilbert space at each point
$\mathbf{k}$ in the 1BZ in a sufficiently large energy range, which
is dubbed as ``disentanglement window'', or ``outer window'',
and then chooses a $J$-dimensional subspace that gives the smallest
possible value of $\varOmega_{\mathrm{I}}$ via iterative procedures.
Similarly to above discussion, the minimization of $\varOmega_{\mathrm{I}}$
also begins with a rough initial guess that are usually obtained by
first projecting $J$ localized trial orbitals $\left\{ g_{n}\left(\mathbf{r}\right)\right\} $
onto $\mathcal{J}_{\mathbf{k}}$ Bloch states such that
\begin{eqnarray}
\left|\phi_{n\mathbf{k}}\right\rangle  & = & \sum_{j=1}^{\mathcal{J}_{\mathbf{k}}}\left|j\,\mathbf{k}\right\rangle \braket{j\,\mathbf{k}|g_{n}},\label{eq:projection}
\end{eqnarray}
and then constructing the following $J$ orthonormalized  Bloch-like states $\widetilde{\left|i\,\mathbf{k}\right\rangle }$
that are related to the original $\left|i\,\mathbf{k}\right\rangle $
via a $\mathcal{J}_{\mathbf{k}}\times J$ matrix $V_{\mathbf{k}}$
such as
\begin{eqnarray}
\widetilde{\left|i\,\mathbf{k}\right\rangle } & = & \sum_{j=1}^{\mathcal{J}_{\mathbf{k}}}\left|j\,\mathbf{k}\right\rangle V_{\mathbf{k},ij}.
\end{eqnarray}
An algebraic  algorithm that updates the minimization iteratively is then used to obtained $J$-dimensional optimal subspace in a self-consistent manner. This process is called subspace selection.
Once the optimal subspace is determined, \textit{i.e.}, the minimization
of $\varOmega_{\mathrm{I}}$ is achieved, the gauge-selection step
to further minimize the noninvariant part $\tilde{\varOmega}$
follows, until a set of self-consistent MLWFs are obtained.

\bibliographystyle{JHEP1}
\addcontentsline{toc}{section}{\refname}\bibliography{Excitation_Semiconductor}

\end{document}